\documentclass[10pt,letterpaper]{article}
\usepackage[margin=1in]{geometry}
\usepackage{times}
\usepackage{graphicx}
\usepackage{amsmath,amssymb}
\usepackage{url}
\usepackage{booktabs}
\usepackage{multirow}
\usepackage{caption}
\usepackage{subcaption}
\usepackage{float}
\usepackage{authblk}
\usepackage{abstract}
\usepackage[ruled,vlined]{algorithm2e}
\usepackage{hyperref}  

\twocolumn 

\title{\textbf{RRRA: Resampling and Reranking through a Retriever Adapter}}

\author[1]{Bongsu Kim}
\affil[ ]{\texttt{sorjdwkd@jbnu.ac.kr}}

\date{}

\begin{document}

\maketitle

\begin{abstract}
In dense retrieval, effective training hinges on selecting high-quality hard negatives while avoiding false negatives. Recent methods apply heuristics based on positive document scores to identify hard negatives, improving both performance and interpretability. However, these global, example-agnostic strategies often miss instance-specific false negatives. To address this, we propose a learnable adapter module that monitors Bi-Encoder representations to estimate the likelihood that a hard negative is actually a false negative. This probability is modeled dynamically and contextually, enabling fine-grained, query-specific judgments.  The predicted scores are used in two downstream components: (1) resampling, where negatives are rewei-ghted during training, and (2) reranking, where top-$k$ retrieved documents are reordered at inference. Empirical results on standard benchmarks show that our adapter-enhanced framework consistently outperforms strong Bi-Encoder baselines, underscoring the benefit of explicit false negative modeling in dense retrieval.
\end{abstract}

\section{Introduction}

Dense retrieval is a core technique in open-domain question answering and document retrieval, using vector-based similarity to match queries with relevant documents~\cite{karpukhin2020dpr}. Its performance largely depends on selecting informative hard negatives—non-relevant documents that are semantically close to the query~\cite{faghri2017vsepp,robinson2020hardnegatives,kalantidis2020hardnegativemix}. These examples provide meaningful gradients, sharpening decision boundaries and improving generalization~\cite{he2020moco,chen2020simclr}.

However, not all hard negatives are beneficial. Some, known as false negatives, are actually relevant but incorrectly labeled as negatives. Using these during training introduces conflicting supervision, distorts the embedding space, and hinders convergence~\cite{schroff2015facenet,chuang2020debiased}. In contrastive learning, this leads to excessive loss and suboptimal optimization~\cite{robinson2020hardnegatives}.

To mine stronger negatives, recent methods adopt self-improving strategies where the model selects top-$k$ candidates based on similarity scores~\cite{xiong2020ance,zhou2022simans}. While this increases difficulty, it also raises the risk of false negatives\-especially under noisy or incomplete labels~\cite{qu2020rocketqa}.

\begin{figure}[t]
  \centering
  \includegraphics[width=\linewidth]{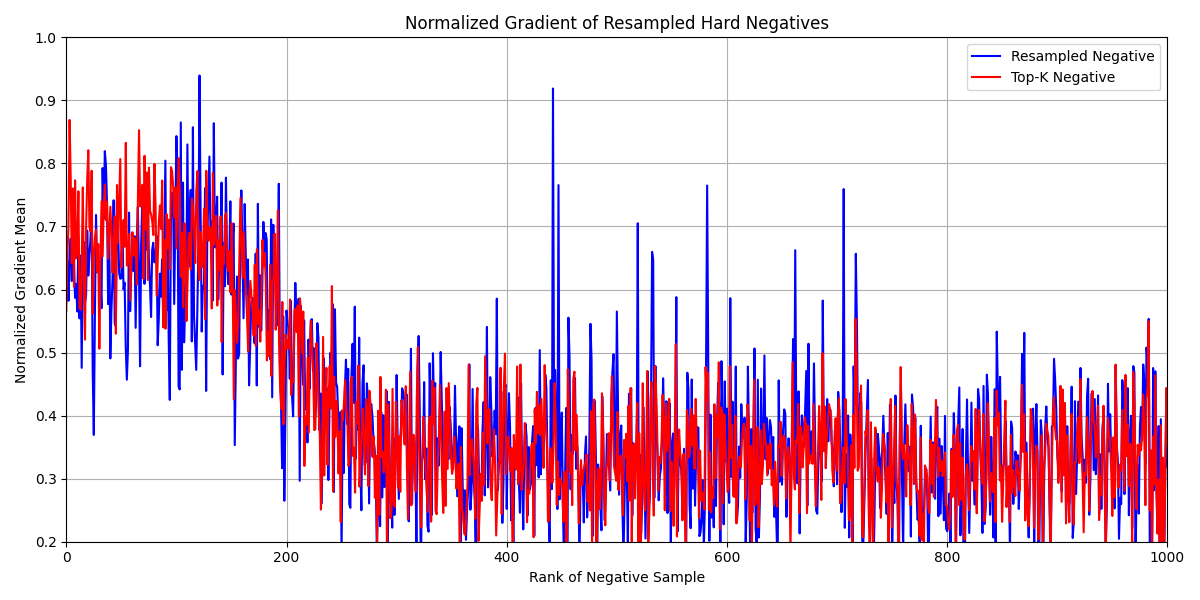}
  \caption{Normalized Gradient of Resampled Hard Negatives}
  \label{fig:motivation}
\end{figure}

To mitigate this, prior work like ADORE~\cite{zhan2021hardnegatives} and SimANS~\cite{zhou2022simans} applies heuristic filters that downsample negatives with high similarity to positives, typically using global thresholds (e.g., mean or variance). Though partially effective, these methods ignore query-specific variation. As models become more performant, they must also capture edge cases where false negatives deviate subtly from true negatives, necessitating finer-grained handling (see Figure~\ref{fig:motivation}). In practice, negative distributions vary across queries, and uniform filtering may discard useful samples or retain harmful ones~\cite{ren2021pair}.

To address this, we propose a learnable adapter that estimates the false negative probability of each candidate by observing intermediate representations from a bi encoder. This probability guides (1) reweighting of negatives during training, and (2) reranking of retrieved documents during inference.

We also conduct gradient-based analyses showing that false negatives exhibit distinguishable patterns from true negatives~\cite{chuang2020debiased}. Our adapter further acts as a lightweight reranker, offering correction signals comparable to cross-encoders~\cite{devlin2019bert,ren2021rocketqav2} while retaining the efficiency of dual encoders~\cite{luan2021sparsedenseattentional}.

Experiments on benchmarks such as DPR show that our method consistently outperforms strong baselines~\cite{karpukhin2020dpr,zhou2022simans,ren2021pair}. The proposed framework is simple, efficient, and broadly applicable, offering an effective solution to the persistent challenge of false negatives in dense retrieval.

\section{Related Work}

\subsection{Retrievers}
Traditional retrieval methods such as TF-IDF and BM25 rely on lexical overlap \cite{yang2017anserini,irbook}, limiting their ability to capture semantic similarity \cite{irbook}. Dense retrieval improves upon this by encoding queries and documents into dense vectors using pretrained language models (PLMs) \cite{gao2021readyfordense}, enabling semantic matching and showing superior performance \cite{dense_survey}. Models like ColBERT alleviate inference cost via late interaction \cite{khattab2020colbert}.

In-batch negatives in DPR simplify training with only positive labels \cite{karpukhin2020dpr}, but offer limited negative diversity \cite{xiong2020ance}. Random negatives often lack difficulty \cite{zhan2021hardnegatives}, while false negatives—relevant documents mislabeled as negatives disrupt contrastive learning \cite{chuang2020debiased}. In MS MARCO, up to 70\% of top-ranked but unlabeled passages are false negatives \cite{qu2020rocketqa}.

\subsection{Reranking Methods}
Cross-encoders provide strong reranking by modeling fine-grained interactions \cite{nogueira2019doctttttquery,ren2021rocketqav2}, but are costly. Joint training approaches (e.g., RocketQAv2 \cite{ren2021rocketqav2}, AR2 \cite{ren2021pair}) integrate retrieval and ranking losses or auxiliary tasks like false negative detection \cite{chuang2020debiased}, yet suffer from increased complexity. SimANS and RocketQA perform soft reranking with dual encoders using adjusted similarity or quality scores \cite{zhou2022simans,qu2020rocketqa}, though these still lack cross-encoder granularity.

\subsection{Negative Sampling}
Negative sampling strategies are central to contrastive learning in domains such as image-text matching and face recognition \cite{faghri2017vsepp,kalantidis2020hardnegativemix,robinson2020hardnegatives,schroff2015facenet}, and in dense retrieval they directly influence representation quality.

Early methods used random or in-batch negatives \cite{karpukhin2020dpr}, which are simple but often uninformative \cite{zhan2021hardnegatives}. ANCE improves this with self-refreshing negatives that evolve during training \cite{xiong2020ance}, and ANCE-Tele further diversifies them via distant sampling \cite{sun2022teleportationnegatives}. However, these still risk including false negatives—semantically relevant but unlabeled documents—which distort the embedding space \cite{chuang2020debiased}. RocketQA mitigates this by filtering with cross-encoder judgments \cite{qu2020rocketqa}, though such supervision is costly and limited to training.

Recent work explores \textbf{learnable samplers}. ADORE learns a query-conditioned sampling distribution updated with gradient feedback \cite{zhan2021hardnegatives}, enabling dynamic, data-driven selection. SimANS models score distributions to filter high-confusion negatives \cite{zhou2022simans}, and TriSampler adds a \textbf{geometric constraint} among query, positive, and negative embeddings \cite{ren2021pair}, improving informativeness while avoiding misleading samples.

These methods reflect a shift from static heuristics to \textbf{context-aware, adaptive sampling}, crucial under label incompleteness. Yet most still rely only on retrieval-stage signals and lack false negative estimation in inference.

\begin{figure*}[t]
  \centering
  \includegraphics[width=\textwidth]{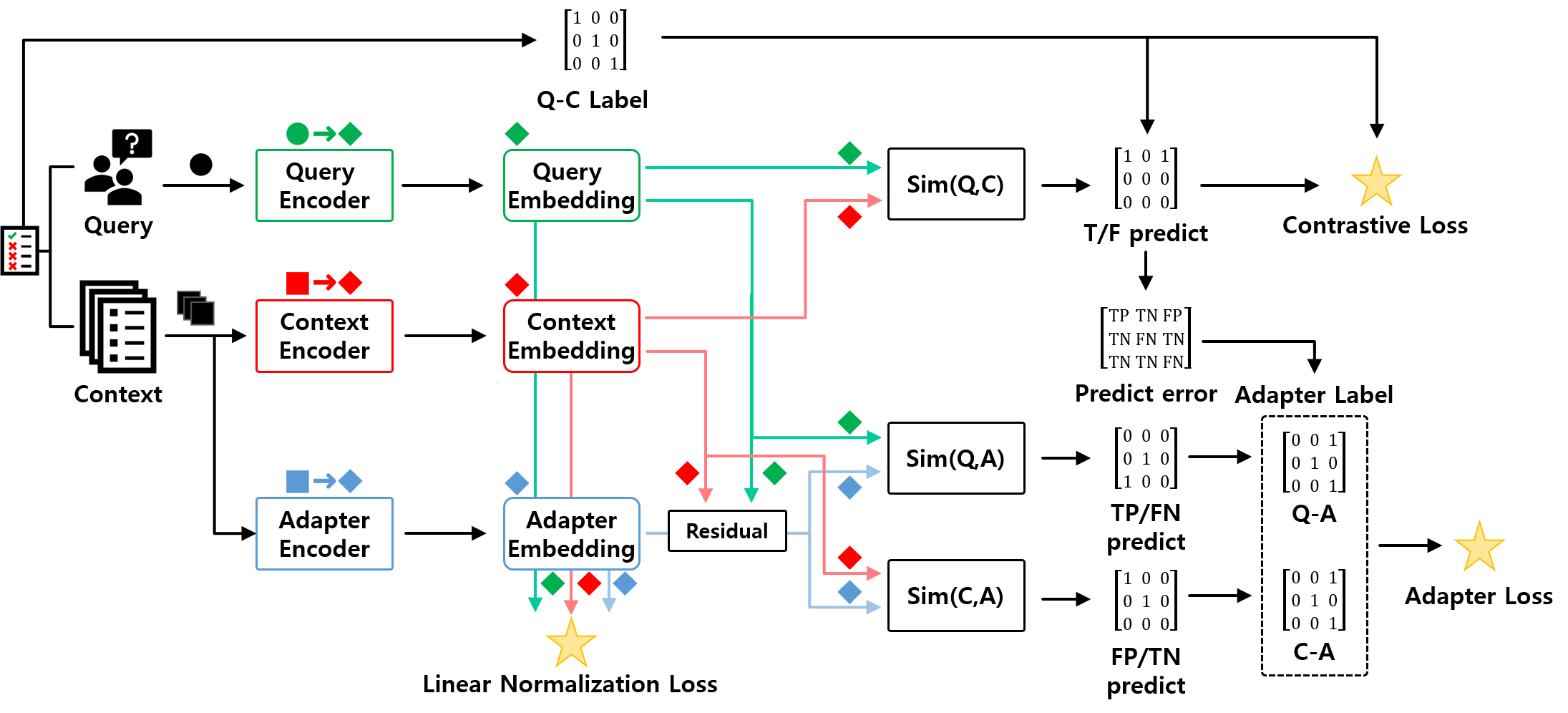}
  \caption{Architecture of the proposed RRRA framework.}
  \label{fig:rrra_archi}
\end{figure*}

\section{Method Overview}

We propose a training framework that addresses the limitations of hard negative sampling in dense retrieval by introducing a mechanism to identify and utilize false negatives. The framework enhances robustness and expressiveness through the following six components:

\begin{itemize}
    \item \textbf{Contrastive Learning with Bi-Encoder:} 
    We start with a standard BERT-based dual encoder trained via contrastive learning to construct a base retrieval space. It pulls positive query–document pairs closer while pushing negatives apart.

    \item \textbf{Adapter-Based Error Detection Task:}
    To detect mislabeled but semantically relevant negatives, we introduce a lightweight adapter module that estimates the false negative likelihood for each candidate based on the query–document pair's joint representation.

    \item \textbf{Adapter–Retriever Integration:}
    To stabilize learning and maintain embedding consistency, the adapter output is added to the original similarity via a residual connection and optionally passed through a normalization layer. This preserves the geometric structure of the encoder’s semantic space.

    \item \textbf{Stage-wise Training Pipeline:}
    Our training follows three phases: (1) pretraining the dual encoder, (2) training the adapter while freezing the encoder, and (3) jointly fine-tuning both to align relevance and correction.

    \item \textbf{Scoring for Re-sampling and Re-ranking:}
    During training, the adapter scores guide reweighting of negative sampling probabilities, suppressing likely false negatives and emphasizing informative examples.
    At inference time, the adapter adjusts top-$k$ retrieval rankings by combining its correction signal with the encoder’s similarity score, improving precision with minimal overhead.
    
\end{itemize}

\section{Representation Framework}

\subsection{Contrastive Learning with Bi-Encoder}

We adopt a BERT-based \textit{Dual Encoder} retriever~\cite{karpukhin2020dense}, where the query encoder \( f_q(\cdot) \) and document encoder \( f_d(\cdot) \) share weights and independently encode:

\[
\mathbf{q} = f_q(q), \quad \mathbf{d} = f_d(d)
\]

Embeddings are derived via average pooling, and similarity is computed as cosine or dot product~\cite{ren2021rocketqa2}:

\[
s(q, d) = \text{sim}(\mathbf{q}, \mathbf{d})
\]

Training uses in-batch negatives for efficient contrastive learning~\cite{xiong2020approximate}. Optimization is performed using AdamW with linear warm-up schedule, consistent with best practices in dense retrieval~\cite{gao2021condenser}. The loss function is binary cross-entropy:

\[
\mathcal{L}_{\text{contrast}} = - \frac{1}{N} \sum_{i=1}^{N} \left[ y_i \log \sigma(s_i) + (1 - y_i) \log\left(1 - \sigma(s_i)\right) \right]
\]

where warm-up and shuffling strategies enhance convergence stability~\cite{he2017decoupled}. This training paradigm forms a strong baseline but lacks cross-document hard negatives, thus motivating our adapter and reranking extensions.


\subsection{Adapter-Based Error Detection Task}

\begin{figure}[t]
  \centering
  \includegraphics[width=\linewidth]{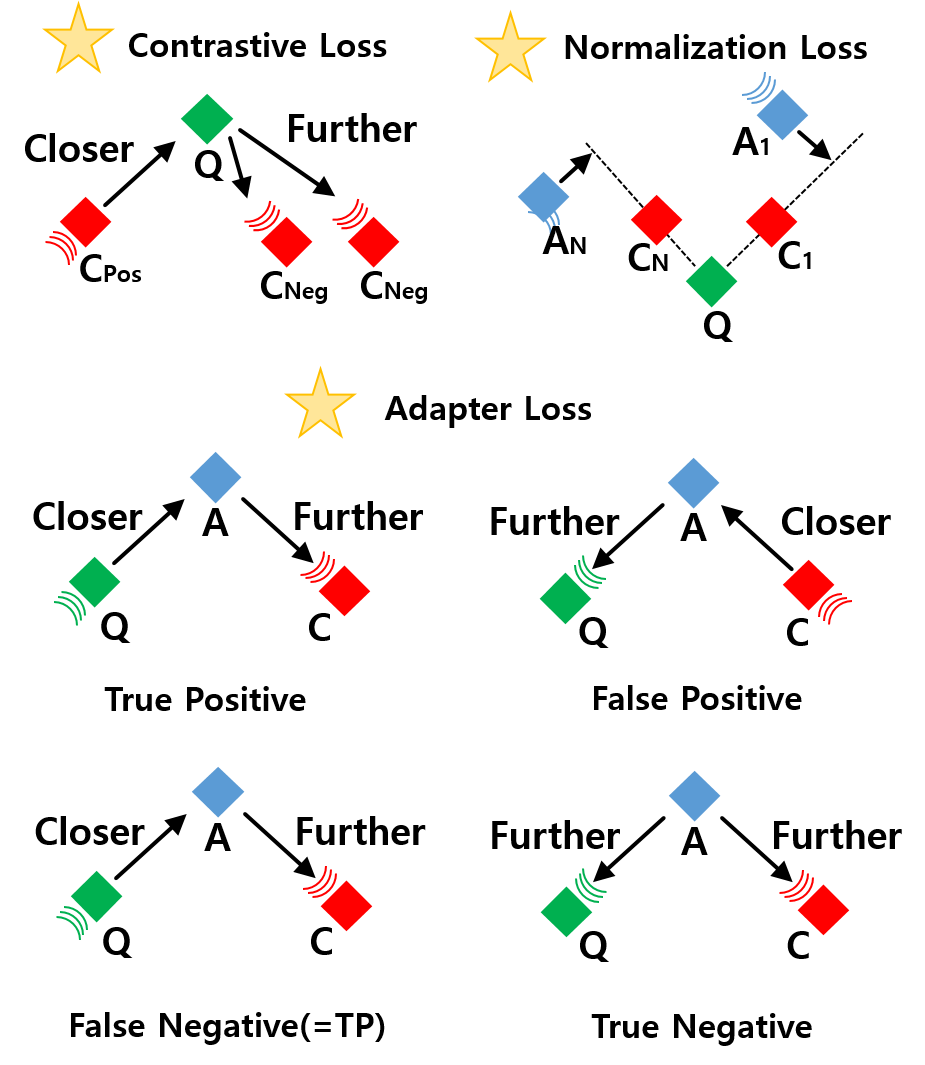}
  \caption{Label distribution predicted by the RRRA adapter.}
  \label{fig:rrra_label}
\end{figure}

The adapter module is trained with two objectives: (1) identifying documents that are semantically similar to positives, and (2) detecting prediction errors made by the dual encoder.

The first objective assumes that false negatives often occur when semantically relevant documents are mistakenly treated as negatives. To capture this, the adapter receives only the context embedding and predicts its ``positive-likeness.'' Initialized from the context encoder, the adapter produces a residual vector \( \Delta \mathbf{d} \), added to the original embedding:

\[
\mathbf{d}_{\text{adapted}} = \mathbf{d} + \Delta \mathbf{d}
\]

This enables soft correction while remaining grounded in the pretrained space.

The second objective is to reclassify predictions into four categories: true positive (TP), false negative (FN), false positive (FP), and true negative (TN)~\cite{yang2024hardness}. These labels provide directional supervision:

\begin{itemize}
  \item TP / FN: move adapted embedding toward the query
  \item FP: move closer to original context
  \item TN: move away from both
\end{itemize}

Given dual encoder prediction \( \hat{y}_{\text{de}} = \sigma(s(\mathbf{q}, \mathbf{d})) \), the adapter outputs the corrected error class. To handle class imbalance, the adapter loss is weighted:

\[
\mathcal{L}_{\text{adapter}} = \frac{1}{N} \sum_{i=1}^{N} w_i \cdot \text{CE}(\hat{\mathbf{y}}_i, \mathbf{y}_i)
\]

This allows the adapter to focus on rare but impactful errors. Once trained, it is reused to guide resampling during training and reranking at inference.


\subsection{Adapter–Retriever Integration}

While the adapter operates on document embeddings alone, this is often insufficient for detecting subtle errors such as false negatives (FN) or false positives (FP). To address this, we introduce (1) a residual correction based on qu\-ery–context interaction, and (2) a geometric constraint to align adapter outputs with the retriever’s semantic space.

\subsubsection{Relation-Aware Residual Correction}

To inject relational cues, we construct an input vector for the adapter using:

\[
\mathbf{z} = \text{concat}(\mathbf{q} - \mathbf{c},\ \mathbf{q} \odot \mathbf{c},\ \mathbf{q} + \mathbf{c})
\]

This captures difference, interaction, and composition between the query and context, inspired by relational modeling in sentence matching tasks~\cite{relationnet, relbert}. A multi-layer perceptron (MLP) maps \( \mathbf{z} \) to a residual correction:

\[
\mathbf{c}' = \mathbf{c} + \Delta \mathbf{c}, \quad \Delta \mathbf{c} = \text{MLP}(\mathbf{z})
\]

The refined embedding \( \mathbf{c}' \) adjusts toward \( \mathbf{q} \) for suspected FNs, remains near \( \mathbf{c} \) for TNs/FPs, or interpolates for ambiguous cases. This lightweight design introduces relational refinement with minimal overhead.

\subsubsection{Linear Normalization Constraint}

To ensure semantic alignment, we constrain the adapted embedding \( \mathbf{a} \) to lie on the line segment between \( \mathbf{q} \) and \( \mathbf{c} \):

\[
\mathbf{a} = \alpha \mathbf{q} + (1 - \alpha) \mathbf{c}, \quad \alpha \in [0, 1]
\]

This positioning makes \( \mathbf{a} \) interpretable: near \( \mathbf{q} \) for positives, \( \mathbf{c} \) for negatives, or intermediate for uncertain cases. We enforce this with a soft regularization:

\[
\mathcal{L}_{\text{norm}} = \frac{1}{N} \sum_{i=1}^N \min_{\alpha \in [0,1]} \left\| \mathbf{a}_i - \left(\alpha \mathbf{q}_i + (1 - \alpha) \mathbf{c}_i \right) \right\|_2^2
\]

This constraint promotes stability, interpretability, and better generalization, especially in identifying false negatives under distributional shift~\cite{normconstraint2022}. It also encourages the adapter to preserve the retriever’s geometry while offering meaningful corrections.


\section{Training and Inference Strategy}

\subsection{Stage-wise Training Pipeline}

Our training strategy is divided into three sequential phases. We first train a dual encoder for base retrieval, then introduce an adapter to detect prediction errors, and finally jointly fine-tune both modules for end-to-end alignment.

\subsubsection{Step 1: Dual Encoder Pretraining}

We begin by training a BERT-based dual encoder using in-batch negatives to align positive query–context pairs and separate irrelevant ones. 
Similarity is computed via dot product, and optimization uses the binary contrastive loss.  
This establishes a base representation space that serves as input to the adapter module.

\subsubsection{Step 2: Adapter Training}

With the encoder frozen, we train the adapter to classify query–context pairs into four categories: true positive (TP), false negative (FN), false positive (FP), or true negative (TN). Training uses:

\begin{itemize}
    \item \textbf{Classification Loss}:
    \[
    \mathcal{L}_{\text{adapter}} = \frac{1}{N} \sum_{i=1}^{N} w_i \cdot \text{CE}(\hat{\mathbf{y}}_i, \mathbf{y}_i)
    \]
    where \(w_i\) is a class-balancing weight and CE denotes 4-way cross-entropy.

    \item \textbf{Normalization Loss}:
    \[
    \mathcal{L}_{\text{norm}} = \frac{1}{N} \sum_{i=1}^N \min_{\alpha \in [0,1]} \left\| \mathbf{a}_i - (\alpha \mathbf{q}_i + (1 - \alpha) \mathbf{c}_i) \right\|_2^2
    \]
    which encourages the adapted embedding \(\mathbf{a}\) to lie along the line between \(\mathbf{q}\) and \(\mathbf{c}\).
\end{itemize}

Adapter scores are later used to resample top-$k$ negatives by estimating their false negative probability.

\subsubsection{Step 3: Joint Fine-tuning}

We jointly fine-tune the bi-encoder and adapter to refine both representations and error signals. 
Negatives are rewe\-ighted using the adapter, and each batch includes a mix of hard and random negatives for stable optimization. 
The total loss combines the contrastive and adapter objectives, omitting the normalization loss to allow greater representational flexibility:

\[
\mathcal{L} = \mathcal{L}_{\text{contrastive}} + \lambda \cdot \mathcal{L}_{\text{adapter}}
\]

This joint training enables mutual correction between modules and improves both ranking and robustness.

\begin{algorithm}[t]
\caption{Step 3: Joint Fine-tuning of Dual Encoder and Adapter}
\label{alg:joint_training}
\KwIn{Pretrained encoders $f_q, f_d$, trained adapter $g_\phi$, adapter-weighted hard negatives}
\KwOut{Fine-tuned $f_q, f_d, g_\phi$}

Enable training for $f_q, f_d, g_\phi$\;

\For{each epoch}{
    Sample mini-batch with adapter-weighted hard negatives and random negatives\;
    \For{each $(q,d)$ in batch}{
        $q \leftarrow f_q(q)$,\quad $d \leftarrow f_d(d)$\;
        $z \leftarrow [q-d,\ q \odot d,\ q+d]$\;
        $\tilde{d} \leftarrow d + g_\phi(z)$\;
    }
    Compute $\mathcal{L}_{\mathrm{contrastive}}$ and $\mathcal{L}_{\mathrm{adapter}}$ (TP/FN/FP/TN)\;
    $\mathcal{L}_{\mathrm{joint}} \leftarrow \mathcal{L}_{\mathrm{contrastive}} + \alpha \mathcal{L}_{\mathrm{adapter}}$\;
    Update $f_q, f_d, g_\phi$ with $\mathcal{L}_{\mathrm{joint}}$\;
}

\textit{Inference: Combine encoder and adapter outputs for reranking.}
\end{algorithm}


\subsection{Scoring for Re-sampling and Re-ranking}

\begin{figure}[t]
  \centering
  \includegraphics[width=\linewidth]{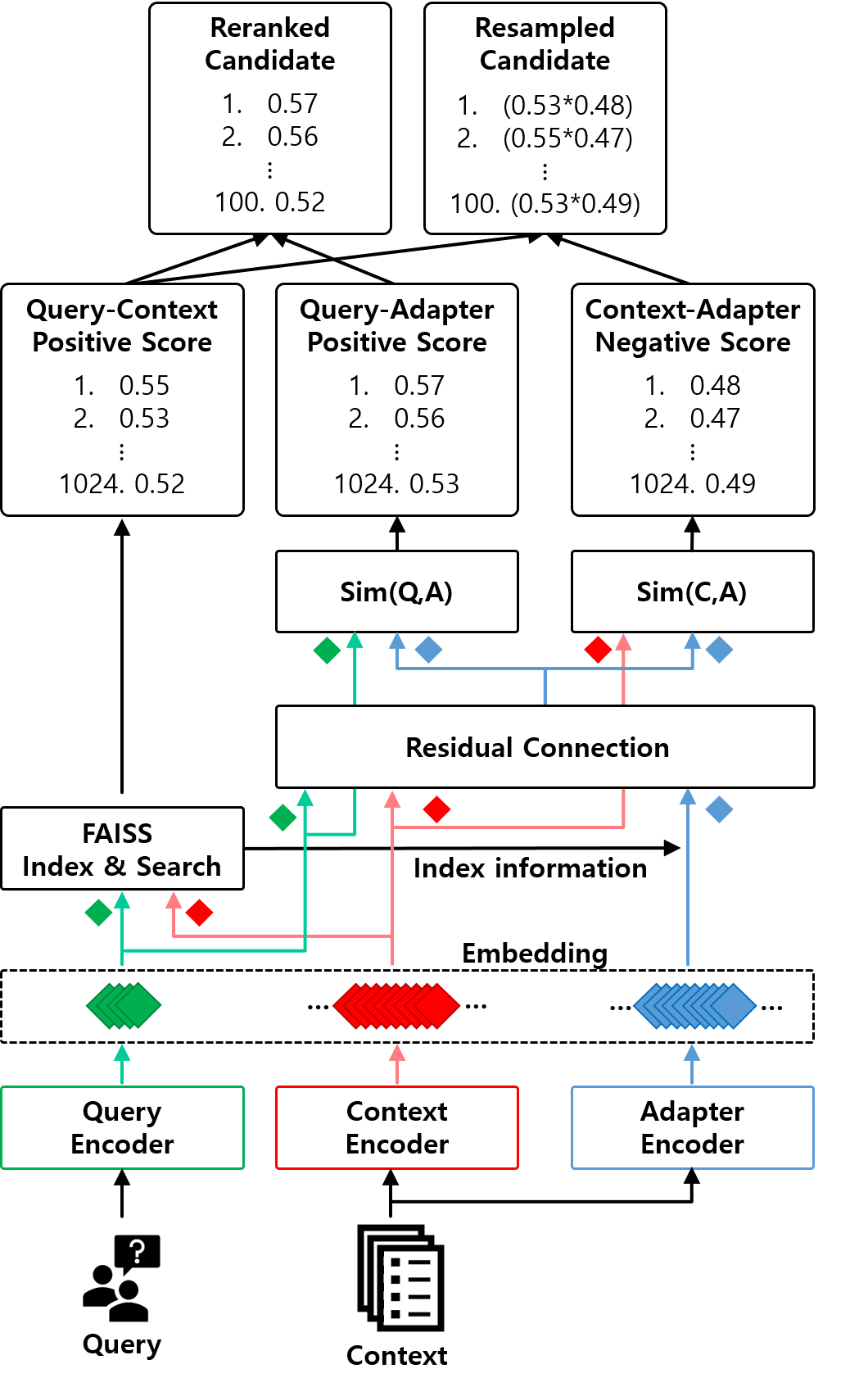}
  \caption{Effect of the ReSampling and ReRanking components.}
  \label{fig:resample_rerank}
\end{figure}

To improve hard negative selection and mitigate the impact of false negatives, we introduce a dual scoring mechanism that evaluates each context candidate with two simil\-arity based scores: one for informativeness and the other for false negative likelihood.

\subsubsection{Score Definitions}

Given a query embedding $\mathbf{q}$, a context embedding $\mathbf{c}$, and its adapter-refined embedding $\mathbf{a}$ , we define:

\begin{itemize}
    \item \textbf{Hard Negative Score} $s_{\text{HN},i}$: semantic similarity between $\mathbf{q}$ and $\mathbf{a}$, indicating informativeness.
    \item \textbf{False Negative Score} $s_{\text{FN},i}$: similarity between $\mathbf{a}$ and $\mathbf{c}$, representing the likelihood that the sample is a false negative.
\end{itemize}

Both scores are computed from the same context text by passing it through the context encoder and adapter.

\subsubsection{Re-sampling Score (Training)}

From the top-$k$ mined negatives (via the frozen bi-encoder), the adapter scores are used to prioritize informative negatives (TP, FN) and down-weight or remove likely false negatives (FP, TN).  
The re-sampling score is computed as:

\[
s_i^{\text{RS}} = s_{\text{HN},i} \cdot \left( 1 - s_{\text{FN},i} \right)^{\gamma_{\text{RS}}}
\]

where $\gamma_{\text{RS}}$ is a hyperparameter controlling the suppression strength for high false-negative likelihood.

\subsubsection{Re-ranking Score (Inference)}

During inference, the adapter provides a lightweight reranking signal.  
Let $s_{\text{Base},i}$ be the base relevance score from the bi-encoder, and $s_{\text{Adapter},i}$ the similarity between $\mathbf{q}$ and $\mathbf{a}$.  
The final reranking score is:

\[
s_i^{\text{RR}} = s_{\text{Base},i} \cdot s_{\text{Adapter},i}^{\lambda_{\text{RR}}}
\]

where $\lambda_{\text{RR}}$ controls the influence of the adapter's correction.

\subsubsection{Practical Considerations}

Because $\mathbf{c}$ and $\mathbf{a}$ can be precomputed and indexed with FAISS, both scores can be computed efficiently in parallel.  
The same scoring framework is shared across training-time re-sampling and inference-time re-ranking, ensuring consistent evaluation of candidates across stages.

By integrating the adapter’s learned false negative discrimination ability with the bi encoder’s base representation, this dual scoring mechanism improves both training robustness and inference accuracy in dense retrieval.


\section{Experiments}

\begin{table*}[t]
\centering
\begin{tabular}{l|ccccc||ccccc}
\toprule
\multirow{2}{*}{\textbf{Method}} &
\multicolumn{5}{c||}{\textbf{NQ}} &
\multicolumn{5}{c}{\textbf{TQ}} \\
 & r@1 & r@10 & r@20 & r@50 & r@100 & r@1 & r@10 & r@20 & r@50 & r@100 \\
\midrule
Bi-Encoder & 51.8 & 73.9 & 81.1 & 84.1 & 86.5 & 57.7 & 77.8 & 79.8 & 82.4 & 85.9 \\
+ Random & 50.3 & 77.1 & 79.6 & 85.5 & 86.6 & 59.0 & 78.0 & 79.0 & 82.0 & 85.5 \\
+ ANCE & 54.8 & 77.3 & 82.5 & 85.8 & 87.9 & 58.8 & 78.2 & 81.4 & 83.2 & 85.9 \\
+ SIMANS & 59.7 & 79.5 & 84.3 & 87.0 & 89.1 & 62.4 & 81.5 & 83.9 & 85.7 & 87.1 \\
+ TriSampler & 59.6 & 79.7 & 84.6 & 87.2 & 89.4 & 62.4 & 81.8 & 84.2 & 86.0 & 87.7 \\
RRRA w/o ReSampling & 58.4 & 78.3 & 82.4 & 85.8 & 88.0 & 58.9 & 80.0 & 83.3 & 84.9 & 86.6 \\
RRRA w/o ReRanking & 63.3 & 79.8 & \textbf{84.7} & 86.9 & \textbf{89.7} & 62.6 & 82.1 & \textbf{84.5} & \textbf{86.2} & \textbf{87.9} \\
RRRA (full) & \textbf{65.9} & \textbf{80.2} & \textbf{84.7} & \textbf{87.0} & 89.6 & \textbf{63.7} & \textbf{81.9} & 84.4 & 86.1 & \textbf{87.9} \\
\bottomrule
\end{tabular}
\caption{Retrieval performance on NQ and TQ datasets. All scores are scaled by 100.}
\label{tab:main_nq_tq}
\end{table*}

\begin{table*}[t]
\centering
\begin{tabular}{l|ccccc||ccccc}
\toprule
\multirow{2}{*}{\textbf{Method}} &
\multicolumn{5}{c||}{\textbf{MS-Pas}} &
\multicolumn{5}{c}{\textbf{MS-Doc}} \\
 & r@1 & r@10 & r@20 & r@50 & r@100 & r@1 & r@10 & r@20 & r@50 & r@100 \\
\midrule
Bi-Encoder & 14.3 & 55.4 & 67.6 & 80.2 & 87.9 & 18.6 & 58.3 & 69.6 & 81.4 & 86.9 \\
+ Random & 14.3 & 55.3 & 67.7 & 80.9 & 89.2 & 16.5 & 60.0 & 71.5 & 82.0 & 87.7 \\
+ ANCE & 14.6 & 54.9 & 67.5 & 81.3 & 89.5 & 16.8 & 61.9 & 74.7 & 85.0 & 90.0 \\
+ SIMANS & 17.4 & 58.0 & 69.2 & 81.9 & 89.9 & 17.7 & 63.8 & 75.3 & 86.4 & 90.7 \\
+ TriSampler & 17.3 & 58.2 & 69.7 & 82.3 & 90.9 & 17.8 & 64.0 & 75.7 & 87.0 & 91.2 \\
RRRA w/o ReSampling & 17.1 & 58.3 & 68.3 & 82.1 & 87.9 & 20.8 & 62.5 & 72.6 & 82.4 & 87.4 \\
RRRA w/o ReRanking & 17.2 & 58.8 & \textbf{69.9} & \textbf{83.8} & \textbf{91.1} & 17.3 & 64.1 & \textbf{76.6} & \textbf{87.5} & \textbf{91.7} \\
RRRA (full) & \textbf{18.8} & \textbf{59.4} & \textbf{69.9} & 83.5 & 90.4 & \textbf{22.4} & \textbf{65.7} & 76.1 & 86.9 & \textbf{91.7} \\

\bottomrule
\end{tabular}
\caption{Retrieval performance on MS-Pas and MS-Doc datasets. All scores are scaled by 100.}
\label{tab:main_mspas_msdoc}
\end{table*}

\subsection{Experimental Setup}

\paragraph{Datasets.}
We evaluate on four standard benchmarks: Natural Questions (NQ)~\cite{kwiatkowski2019nq}, TriviaQA (TQA)~\cite{joshi2017triviaqa}, MS \-MARCO Passage (MSPas), and MS MARCO Document (MSDoc)\allowbreak~\cite{nguyen2016msmarco}. Following prior work~\cite{zhou2022simans,ren2021pair}, we sample negatives using a 1:15 positive-to-negative ratio. For each query, we retrieve the top-1024 passages with Faiss~\cite{johnson2019billionscale} as negative candidates.

\begin{table}[H]
\centering
\begin{tabular}{lrrrr}
\toprule
\textbf{Dataset} & \textbf{Train} & \textbf{Dev} & \textbf{Test} & \textbf{\# Documents} \\
\midrule
NQ     & 58{,}880  & 8{,}757  & 3{,}610  & 21{,}015{,}324 \\
TQA    & 60{,}413  & 8{,}837  & 11{,}313 & 21{,}015{,}324 \\
MSPas  & 502{,}939 & 6{,}980  & -        & 8{,}841{,}823  \\
MSDoc  & 367{,}013 & 5{,}193  & -        & 3{,}213{,}835  \\
\bottomrule
\end{tabular}
\caption{Statistics of datasets used for training and evaluation.}
\label{tab:dataset_stats}
\end{table}

\paragraph{Evaluation Metrics.}
We use Recall@$k$ ($k \in \{1,5,10,\allowbreak20,50,100\}$) to evaluate both retrieval depth and top ran\-ked precision. R@1 and R@5 are especially informative in reranking scenarios.

\paragraph{Baselines.}
We use BERT-base as the backbone and first evaluate bi-encoder performance. We then compare against random sampling, ANCE~\cite{xiong2020ance}, SimANS~\cite{zhou2022simans}, and TriSampler~\cite{ren2021pair}, all using the same pretrained checkpoint. Our method (RRRA) is tested under two configurations: ReSampling, where the adapter reorders hard negatives during training, and ReRanking, where it reranks candidates at inference. We also report results from each stage separately and include comparisons with external baselines such as BM25~\cite{yang2017anserini}, DPR~\cite{karpukhin2020dpr}, RocketQA~\cite{qu2020rocketqa}, and ADORE\allowbreak~\cite{zhan2021hardnegatives}.

\paragraph{RRRA Details.}
We adopt a BERT-base dual encoder as the backbone across all stages. 
In Stage~1 and Stage~2, training is performed with a batch size of 128. 
Stage~3 uses a batch size of 64 with gradient accumulation steps of 2 (effective batch size 128) and incorporates 4 hard negatives per query in each batch. 
For adapter training in Stage~2, the \textit{ContextE Init} option initializes the adapter from the context encoder before learning. 

Hyperparameters are set as follows: the class imbalance weighting parameter $\gamma_{\text{imb}}$ in Stage~2 is $0.3$; the resampling suppression factor $\gamma_{\text{RS}}$ and the reranking adjustment factor $\lambda_{\text{RR}}$ are tuned separately on the development set.

\subsection{Results on NQ and TQ}

Table~\ref{tab:main_nq_tq} shows retrieval performance on NQ and TQ. RR\-RA achieves the best results across all ranks, with 65.9 (R@1) and 89.6 (R@100) on NQ, and 63.7 and 87.9 on TQ. Compared to SimANS, RRRA improves R@1 by +6.2 on NQ and +1.2 on TQ.

We evaluate RRRA in three configurations: with only reranking, only resampling, and with both. Reranking (RRRA w/o ReSampling) yields better performance at top ranks such as R@1 and R@10, while resampling (RRRA w/o ReRanking) shows greater gains at deeper ranks like R@50 and R@100. The full model, combining both, consistently outperforms all baselines.

These results show that reranking enhances top-ranked precision, while resampling improves training quality. Despite using only a bi-encoder and a lightweight adapter, RRRA surpasses strong baselines including ANCE, Sim\-ANS, and TriSampler, significantly outperforming heuristic methods.

\subsection{Results on MS MARCO}

Table~\ref{tab:main_mspas_msdoc} presents results on MS-Pas (passage-level) and MS-Doc (document-level). RRRA consistently outperforms baselines, including SimANS and TriSampler, showing clear gains across ranks.

On MS-Pas, RRRA achieves 18.8 (R@1) and 90.4 (R@100), improving over SimANS by +1.4 and +0.5, particularly at top ranks (R@1, R@10, R@20), where reranking is most effective.

On MS-Doc, RRRA achieves 22.4 (R@1) and 91.7 (R@100), outperforming SimANS by +4.7 and +1.0. This confirms its robustness even in document-level settings with longer inputs.

As with NQ and TQ, reranking (RR\-RA w/o ReSampling) improves top-ranked recall, while resampling (RRRA w/o ReRanking) is more effective at deeper ranks. Removing reranking causes a larger drop at R@1 (-5.1 on MS-Doc), while removing resampling affects R@100 less severely.

Overall, RRRA delivers consistent gains across MS MARCO datasets, confirming its effectiveness in both fine-grained and large-scale retrieval.

\begin{table}[t]
\centering
\resizebox{\linewidth}{!}{
\begin{tabular}{lcccc}
\toprule
\textbf{Method} & \shortstack{\textbf{NQ}\\\textbf{(R@100)}} & \shortstack{\textbf{TQ}\\\textbf{(R@100)}} & \shortstack{\textbf{MS-Pas}\\\textbf{(R@50)}} & \shortstack{\textbf{MS-Doc}\\\textbf{(R@100)}} \\
\midrule
BM25~\cite{bm25}                 & 73.7 & 76.7 & 59.2 & 80.7 \\
DPR~\cite{dpr}                   & 85.4 & 84.9 & --   & 86.4 \\
ANCE~\cite{ance}                 & 87.5 & 85.3 & 81.1 & 89.4 \\
RocketQA~\cite{rocketqa}         & 88.5 & --   & 85.5 & -- \\
ADORE~\cite{hardnegatives}       & --   & --   & --   & 91.9 \\
SimANS~\cite{simans}             & 90.3 & 88.1 & 88.7 & 92.3 \\
TriSampler~\cite{trisampler}     & 90.7 & 88.5 & 89.1 & 93.1 \\
\midrule
RRRA w/o ReRanking               & 88.0 & 86.6 & 82.1 & 87.4 \\
RRRA w/o ReSampling              & 89.7 & 87.9 & 83.8 & 91.7 \\
RRRA (full)                      & 89.6 & 87.9 & 83.5 & 91.7 \\
\bottomrule
\end{tabular}
}
\caption{Retrieval performance comparison across datasets. Metrics per dataset are reported at standard cutoff ranks.}
\label{tab:compact_baseline_comparison}
\end{table}

\subsection{Comparison with Strong Baselines}

Table~\ref{tab:compact_baseline_comparison} compares RRRA with classical and recent dense retrieval methods on four datasets. We report R@100 for NQ, TQ, and MS-Doc, and R@50 for MS-Pas.

On MS-Pas and MS-Doc, RRRA scores 83.5 and 91.7, close to SimANS and TriSampler. ADORE, which uses a RoBERTa-based encoder, achieves slightly higher performance on MS-Doc, but RRRA shows nearly comparable results with a simpler setup.

TriSampler and SimANS are based on AR2~\cite{ar2}, whi\-ch uses cross-encoder distillation and self-mining. RocketQA\allowbreak~\cite{rocketqa} also adopts staged training and cross-batch negatives. In contrast, RRRA uses a BERT-base encoder and simple contrastive learning.

Ablations show that reranking has the most impact ($-4.3$ on MS-Doc), while resampling offers smaller but consistent gains. Overall, RRRA demonstrates the potential to deliver competitive performance across benchmarks.

\subsection{Ablation Study on Adapter Components}
\begin{table}[t]
\centering
\begin{tabular}{lcc}
\toprule
\textbf{Method} & \textbf{F1 Score (\%)} \\
\midrule
Adapter w/o Residual          & 63.9 \\
Adapter w/o Linear Norm       & 85.2 \\
Adapter w/o FT-FN Ratio       & 90.9 \\
Adapter w/o ContextE Init     & 92.2 \\
\textbf{Full Adapter}         & \textbf{93.3} \\
\bottomrule
\end{tabular}
\caption{Ablation Study on Adapter Components. The F1 score drops as each module is removed, indicating its contribution to overall performance.}
\label{tab:adapter_ablation}
\end{table}

To assess each component's contribution, we conduct an ablation study by selectively disabling individual adapter modules and evaluating F1 score, which measures how accurately the adapter identifies false positives and false negatives during Stage 2.
Table~\ref{tab:adapter_ablation} shows that removing any component leads to performance degradation, confirming the importance of residual connections, normalization, ratio modeling, and initialization (ContextE Init: initializing the adapter from the context encoder before training).
Experiments are conducted on the MS MARCO Document dataset with the full training set, using 5 training epochs and imbalance weighting parameter $\gamma = 0.3$. All components contribute meaningfully to the adapter's effectiveness.

\subsection{Gradient Analysis of Reweighted Negatives}

We analyze Stage 2 supervision by measuring gradients during Stage 3 training, focusing on how negative sampling strategies affect the document encoder. Using the MS MARCO Document dataset, we compare random sampling, top-$k$ mining, SimANS, TriSampler, and RRRA resampling. Gradients are computed over 100 queries with 1,000 candidates each (100,000 pairs total) via forward passes without parameter updates.

RRRA produces well-regulated gradients, maintaining high magnitude near top ranks with low variance. Uninformative negatives tend to retain rank order, while hard or false negatives shift depending on sampling. RRRA resampling yields lower gradients than top-$k$ within rank 200, but higher beyond, indicating effective repositioning of false negatives.

At ranks 0–3, RRRA negatives (blue) fall within 0.55
–0.65 gradient magnitude, compared to 0.65–0.85 for top-$k$ (red), suggesting finer control and reduced noise. These adjustments to negative weights contribute to improved optimization stability and retrieval performance.

\section{Conclusion and Discussion}

We introduced \textbf{RRRA}, a dense retrieval framework that mitigates false negatives through a learnable adapter applied in both training-time resampling and inference-time reranking. 
Unlike heuristic methods using global statistics, RRRA performs \emph{instance-level} modeling from encoder states, enabling targeted filtering, reweighting, and reranking of top-$k$ hard negatives.

By jointly observing query and context encoders, RRR\-A estimates false-negative likelihood and refines context representations via contrastive supervision. 
This allows adaptive resampling of informative negatives during training and lightweight reranking during inference, improving retrieval precision with minimal overhead.
Gradient-based analysis shows that RRRA produces appropriately difficult negatives—maintaining high magnitude with low variance at top ranks—enhancing optimization stability.

Experiments on four benchmarks (NQ, TQ, MS-Pas, MS-Doc) show consistent gains in both R@1 and R@100. 
Reranking yields the largest top-rank improvements, while resampling provides smaller but consistent gains at deeper ranks.
Despite relying on a bi-encoder backbone and light\-weight adapter, RRRA matches or approaches the performance of more complex systems such as SimANS, TriSampler, and ADORE.

A limitation is reliance on the base encoder’s capacity; stronger backbones may further amplify benefits.
Incorporating cross-encoder distillation, as in AR2, is a promising extension to enhance discrimination ability.
Compared to our directly implemented baselines, RRRA shows clear advantages in learning-based false negative filtering and strong potential when further leveraged for reranking.

Overall, RRRA offers a simple, scalable, and modular approach to learning-based negative sampling in dense retrieval, achieving competitive performance without cross-encoder scoring while enabling instance-aware, contrastive representation learning.

\bibliographystyle{unsrt}
\bibliography{refs}

\end{document}